# Crystal Growth and Magneto Transport of $Bi_2Te_3$ topological insulator

Rabia Sultana[1,2], P. Neha[3], R. Goyal[1,2], S. Patnaik[3] and V.P.S. Awana[1,*]

[1]Quantum Phenomena and Applications Division, National Physical Laboratory (CSIR)
Dr. K. S. Krishnan Road, New Delhi-110012, India.
[2]Academy of Scientific and Innovative Research, NPL, New Delhi-110012, India
[3]School of Physical Sciences, Jawaharlal Nehru University, New Delhi, India-110067

We report synthesis, structural details and electrical transport properties of topological insulator $Bi_2Te_3$. The single crystalline specimens of $Bi_2Te_3$ are obtained from high temperature (950°C) melt and slow cooling (2°C/hour). The resultant crystals were shiny, one piece (few cm) and of bright silver color. The $Bi_2Te_3$ crystal is found to be perfect with clear [00l] alignment. The powder XRD pattern being carried out on crushed crystals showed that $Bi_2Te_3$ crystallized in $R\bar{3}m$ symmetry with $a = b = 4.3866(2)$ Å, $c = 30.4978(13)$ Å and $\gamma = 120^0$. The Bi position is refined to (0, 0, 0.4038 (9)) at Wyckoff position $6c$ and of Te are (0, 0, 0) at Wyckoff position $3a$ and at (0, 0, 0.2039(8)) at $6c$. Ambient pressure and low temperature (down to 2K) electrical transport measurements revealed metallic behavior. Magneto transport measurements under magnetic field showed huge non saturating magneto resistance (MR) reaching up to 250% at 2.5K and under 50KOe field. Summarily, the short communication clearly demonstrates that $Bi_2Te_3$ topological insulator exhibit non-saturating large positive MR at low temperature of say below 10K. The non saturating MR is seen right up to room temperature albeit with much decreased magnitude. Worth mentioning is the fact that these crystals are bulk in nature and hence the anomalous MR is clearly an intrinsic property and not due to the size effect as reported for nano-wires or thin films of the same.

*Corresponding Author

Dr. V. P. S. Awana:  E-mail: awana@nplindia.org
Ph. +91-11-45609357, Fax-+91-11-45609310
Homepage: awanavps.webs.com

# INTRODUCTION

Topological insulators (TI) are among the newest wonders in condensed matter. In fact, the ongoing research on TI has attracted huge attention of condensed matter physics community in recent years [1-3]. Particularly, the insulating band gaps in the bulk, but the gap less high conductivity at surface or edge states had been of tremendous interest [4-7]. Basically, the spin of the electron gets locked with the momentum through spin orbit interaction [3]. The topological insulators had been the source of new exotic physics, whereby a band insulator exhibits good conductivity due the topological surface states and the role of the both spin and momentum is found be important [1-7].

Interestingly, enough it was soon realized in early days itself that superconducting state can also be envisaged in a topological insulator and various proposals were put forward for accommodation of superconductivity in topological insulators [8-12]. Precisely echoing the fact the gap less surface/edge conducting states could eventually be the superconducting one as well [8]. Namely to cite a few superconductivity is reported in $(Cu/Sr)_x Bi_2Se_3$ [13-16], $Tl_{0.6}Bi_2Te_3$ [17], $Nb_xBi_2Se_3$ [18], and $Pd_xBi_2Te_3$ [19] etc.

As far as the ground state pristine topological insulators i.e., $Bi_2Te_3$, $Bi_2Se_3$ and $Bi_2Sb_3$ etc are concerned, as mentioned above their conduction process is a matter of debate. Principally, because these conduct electricity on their surface via special surface electronic states, which in turn are topologically, protected i.e., these surfaces cannot be damaged by impurities. One of the most studied such topological insulator is $Bi_2Te_3$. $Bi_2Te_3$ is a layered topological insulator well known for its thermoelectric properties and used for several applications such as room- temperature power generation and refrigeration [20]. Also, $Bi_2Te_3$ is a narrow gap semiconductor with an indirect gap of approximately 0.15eV and has a strong spin-orbit interaction [21]. The strong spin orbit interaction gives rise to the inversion of the band structures around the $\Gamma$ point of the Brillouin zone (BZ) analogous to the HgTe quantum wells and thus the topological insulating behavior [22]. On the other hand the presence of large bulk gap in $Bi_2Te_3$ makes it suitable for high- temperature spintronics applications [1-3]. $Bi_2Te_3$ possesses conductive surface state with clear presence of a Dirac cone in its electronic structure [23, 24]. Magneto resistivity of topological insulators is reported by various groups in single crystal nano-wires [25], thin films [26, 27], nano-ribbons [28] and doped crystals [29].

Interestingly, because the topological insulators are suitable for spintronics applications, their magneto transport studies do become interesting and desirable. In fact, non saturating magneto-resistance (MR) is reported [25-29] in various TI at low temperatures below say 10K. However, the role of size (material size close to mean free path) say in case of nano-wires/ribbons [25, 28] and of grain boundaries for thin films [26,27] may affect the intrinsic outcome. On the other hand the role of doping in case of single crystals [29] and ensuing disorder could also limit the intrinsic outcome. Keeping in view the fact that magneto transport studies of TI are much desirable due to their spintronic nature and the necessity of the clean data on pristine single crystalline TI, here in this letter we report the same for pristine $Bi_2Te_3$ single crystal in applied field of up to 5 Tesla and down to 2.5K.

**EXPERIMENTAL DETAILS**

Sample of compositions $Bi_2Te_3$ was grown in single crystalline form by self flux from high temperature (950˚C) melt and slow cooling (2˚C /hour). Typically the nominal amounts of 4N pure Bi, and Ti are weighed accurately and mixed thoroughly with the help of mortar and pestle in an Argon filled Glove box. The obtained mixed powder was pressed into a rectangular pellet using the hydraulic press under an approximate pressure of 40Kg/cm$^2$ and then vacuum sealed (10$^{-3}$ Torr) in a quartz tube. The encapsulated quartz tube was then placed in a box furnace and heated to 950˚C (2˚C/min) and kept there for 12 hours. The furnace was then cooled slowly (2˚C/hour) to 650˚C, after which the furnace was allowed to cool down slowly to room temperature (2˚C/min). The schematic diagram of heat treatment is given in Fig. 1(a). The sintered samples obtained were in the form of one piece single crystal, shiny and of bright silver color, see Fig. 1(b). The structural analysis for checking the phase purity was done through room temperature X-ray diffraction (XRD) using Rigaku X-ray diffractometer with Cu-Kα radiation (λ=1.5418 Å). The morphology and compositional analysis of the obtained single crystal is seen by scanning electron microscopy (SEM) images on a ZEISS-EVO MA-10 scanning electron microscope having coupled Energy Dispersive X-ray spectroscopy (EDAX). Raman spectra of bulk $Bi_2Te_3$ single crystal is taken at room temperature using the Renishaw Raman Spectrometer. The electrical transport measurements with and without magnetic field were carried out using Physical Property Measurement System (PPMS).

**RESULTS AND DISCUSSION**

The synthesized $Bi_2Te_3$ compound is crystallized in rhombohedral crystal structure with $R\bar{3}m$ (D5) space group. Also, the crystal structure belongs to the tetradymite group. The presence of rhombohedral structure confirms the topological insulator phase. The (00l) planes are clearly visible from the single crystal XRD pattern shown in Fig. 2(a), a piece of crystal used for surface XRD is shown in inset of Fig. 2(b). The obtained lattice parameters from the Rietveld refinement (Fig. 2b) are $a = 4.3866(2)$ Å, $b = 4.3866(2)$ Å and $c = 30.4978(13)$ Å whereas the values of α, β and γ are 90˚, 90˚ and 120˚. These parameters are in general agreement with earlier reports. The positions of the Bi atoms was refined to Bi (0, 0, 0.4038(9)) at Wyckoff position 6c and the positions of Te atoms were (0, 0, 0) at Wyckoff position 3a and (0, 0, 0.2039(8)) at Wyckoff position 6c. Thus, all the positions refer to the hexagonal setting of the rhombohedral cell. The unit cell is drawn from the observed co-ordinate positions and the lattice parameters, which is shown in Fig. 2(c). It is clear from Fig. 2(c) that the studied $Bi_2Te_3$ crystal exhibits a layered structure composed of quintuple layers (QLs) similar to the $Bi_2Se_3$ compound. Each quintuple layer contains five atomic layers arranged in Te-Bi-Te-Bi-Te fashion which are further bonded by ionic-covalent bonds. However, each quintuple layer is a reverse image of its adjacent quintuple layer and separated by weak van der Waals forces. The presence of van der Waals gap in between the quintuple layers makes the compound easily cleavable [1-3]. Also, intercalation of some doping materials into these van der Waals gap results in the achievement of superconductivity [13-19]. The layered structure of the synthesized compound is confirmed from the SEM images, to be discussed in next section.

Typical scanning electron microscope (SEM) image for the obtained $Bi_2Te_3$ crystal is shown in Fig. 3. The image shows clearly the layered directional growth, which is evidenced by the surface XRD shown in Fig. 2(a). For quantitative elemental analysis the Energy Dispersive X-ray spectroscopy (EDAX) is done on a portion of the SEM image of crystal and the results are shown in insets of Fig. 3. It is clear from inset (a) that except the compositional constituents Bi and Te no other elements including surface contamination from C or O are seen. The quantitative weight% values of summed Bi and Te are shown in inset (b). The calculated formula of the crystal thus comes out to be close to $Bi_2Te_3$. Clearly, the SEM results shown in Fig. 3(a-c) do confirm that the $Bi_2Te_3$ crystal does grown in slab like layered structure along 00l direction and the composition of the crystal is close to the nominal i.e. $Bi_2Te_3$.

Figure 4 illustrates the Raman spectra of bulk $Bi_2Te_3$ single crystal done at room temperature using the Renishaw Raman Spectrometer. Here, the spectra were examined using a laser source having excitation photon energy of 2.41eV (514nm). The measured spectral range is 50–400 $cm^{-1}$ and the spectral resolution is 0.5 $cm^{-1}$ using the 2400 l/mm grating. Apparently, three different Raman vibrational modes namely $A_{1g}^1$, $E_g^2$ and $A_{1g}^2$ are observed. The $A_{1g}^1$ and $A_{1g}^2$ vibrational modes are observed at about 60.74 $cm^{-1}$ and 132.79 $cm^{-1}$ respectively whereas, the $E_g^2$ appears at 102 $cm^{-1}$ which are comparable to the earlier reported result [30].

Fig. 5(a) depicts the resistivity versus temperature ($\rho$-T) plot for the studied $Bi_2Te_3$ crystal in temperature range of 300K down to 2.5K. The $\rho$-T behavior is metallic and the residual resistivity i.e., $\rho$ at 0K is about 10$\mu\Omega$-cm. The room temperature resistivity i.e., $\rho$ (300K) is around 0.18m$\Omega$-cm. These values are comparable to those as being reported earlier for the bulk $Bi_2Te_3$ crystals. Apparently, the value of resistivity decreases with decreasing temperature from 300K. Thus, we can say that it has positive temperature coefficient value and the nature of the curve can be considered as metallic. Here, we can see that resistivity data follows the equation $\rho=\rho_o+AT^2$ in a temperature range from 5K to 50K, which is hallmark of Fermi liquid behavior. The residual resistivity i.e. $\rho$ at 0K by extrapolation comes out to be 0.0114 m$\Omega$-cm. Consequently, the value of residual resistivity ratio (ratio of resistivity at room temperature and at zero temperature) is estimated to be 15.6.

Figure 5(b) shows the temperature dependent resistivity under different magnetic field in the temperature range from 5K to 50K. Clearly, external magnetic field enhances the value of resistivity. All the resistivity curves obey Fermi-liquid behavior as shown in figure 5(a). It is observed that the value of residual resistivity ($\rho_0$) increases from 0.0114 m$\Omega$-cm in zero field to 0.2346 m$\Omega$-cm in 6 Tesla field.

Topological insulators having huge magneto resistance (MR) plays an important role both in basic research as well as from application point of view [26]. The magneto resistance can be described as the tendency of a material to change the value of its electrical resistance in an externally applied magnetic field. Figure 5(c) shows the percentage change of magneto resistance (MR) with respect to applied magnetic field (up to 5Tesla) at different temperatures (from 2.5K to 280K). Here, the value of MR is calculated using the equation MR = [$\rho$(H)- $\rho$(0)]/ $\rho$(0). Fig. 5(c) clearly shows that the MR increases as the applied magnetic field increases but decreases with increase in temperature. For a maximum field of about 5Tesla, the approximate values of MR (%) are 250, 140, 115, 65 and 5 at a temperature of 2.5K, 5K, 10K, 50K and 280K respectively. From the above mentioned values one can clearly observe a large non saturating MR (%) value reaching up to 250% at 2.5K under an applied magnetic field of about 5Tesla and this value is in agreement to those as reported earlier [31]. Similar behavior i.e., origin of giant linear MR theory was proposed in case of doped silver chalcogenides by Abrikosov popularly known as the linear quantum MR theory [32]. However, the linearity of the MR is somewhat suppressed as the temperature is increased.

Figure 5(d) shows the temperature dependent MR ($\rho$(H) - $\rho$(0)/ $\rho$(0)) under different temperature. Clearly, MR data can be best fitted to power law dependence, $\Delta\rho/ \rho(0) \propto B^\alpha$, shown as solid curve. Here, the value of $\alpha$ is closer to 1 in a temperature range from 2.5K to 50K. While, at 280K, the value of $\alpha$ is closer to 1.9, which implies non linear behavior of magneto-resistance, as shown in its inset.

The well established Kohler's rule is used to analyze temperature and field dependent magneto-resistance of many metals. Here, Kohler's rule for a standard metal is defined as the variation in temperature dependent resistivity in an applied magnetic field obeying the functional relation $\Delta\rho/\rho(0) = F(H/\rho_0)$, where $\rho_0$ is the zero resistivity at given temperature [33]. For our $Bi_2Te_3$ single crystal Kohler's Plot is shown in Fig. 5(e). Apparently, the curves at different temperatures are not merging on each other. Kohler's rule says that plot of $\Delta\rho/\rho_o$ as a function of magnetic field at distinct temperatures should merge onto a single curve if there exist single type of charge carriers and the resulting scattering time under magnetic field is same at all points on the Fermi surface [33]. Here, the Kohler plot of the magneto-resistance of $Bi_2Te_3$ does not collapse onto a single curve, therefore we can say that it supports the multi-carriers scattering mechanism, alike quasi one dimensional chain containing compounds [34] and as well as other reports on topological insulators [1,2,35]. There are various suggestions about the transport mechanism in topological insulators including quantum transport [1,2].

Summarily, we have synthesized quality single crystals of $Bi_2Te_3$ topological insulator, of which structural, micro-structural, Raman spectroscopy and electrical transport down to low temperatures (2.5K) under magnetic field (6Tesla) are reported. The as grown $Bi_2Te_3$ crystals are large enough (cm) in size, grown in *c*-direction and do exhibit anomalous non saturating high MR of above 250% at 2K under 6Tesla magnetic field. Worth mentioning is the fact that these crystals are bulk in nature and hence the anomalous MR is clearly an intrinsic property and not due to the size effects as reported for nano-wires or thin films of topological insulators [25-28].

## ACKNOWLEDGENT

Authors from CSIR-NPL would like to thank their Director NPL India for his keen interest in the present work. Authors further thank Dr. Bhasker Gahtori for SEM and Mrs. Shaveta Sharma for Raman studies. This work is financially supported by DAE-SRC outstanding investigator award scheme on search for new superconductors. Rabia Sultana thanks CSIR, India for research fellowship and AcSIR-NPL for Ph.D. registration.

# FIGURE CAPTIONS

**Figure 1(a):** Schematic diagram of heat treatment to grow bulk $Bi_2Te_3$ single crystal.

**Figure 1(b):** Photograph of as grown $Bi_2Te_3$ single crystal.

**Figure 2(a):** X-ray diffraction pattern for $Bi_2Te_3$ single crystal, inset shows the piece of as grown crystal.

**Figure 2(b):** Rietveld fitted room temperature X-ray diffraction pattern for powder $Bi_2Te_3$ crystal.

**Figure 2(c):** Unit cell structure of $Bi_2Te_3$ single crystal.

**Figure 3:** Scanning electron microscopy image for $Bi_2Te_3$ single crystal.

**Figure 4:** Raman Spectra for $Bi_2Te_3$ single crystal at room temperature.

**Figure 5(a):** Temperature dependent electrical resistivity of $Bi_2Te_3$ in a temperature range of 300K to 2.5K. Red solid curve shows Fermi-liquid behavior fitted using equation $\rho=\rho_o+AT^2$

**Figure 5(b):** Temperature dependent electrical resistivity for $Bi_2Te_3$ single crystal under different applied magnetic field.

**Figure 5(c):** MR (%) as a function of magnetic field for $Bi_2Te_3$ at different temperatures.

**Figure 5(d):** Magnetic field dependent magneto-resistance at different temperatures. Solid lines are the fitting results using power law dependence.

**Figure 5(e):** Kohler plot for $Bi_2Te_3$ in a field range from 0 to 5Tesla at several temperatures.

Fig. 1(a)

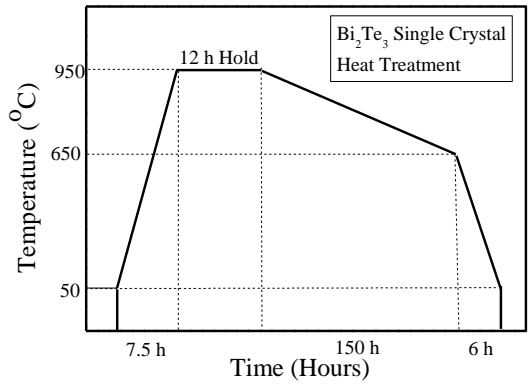

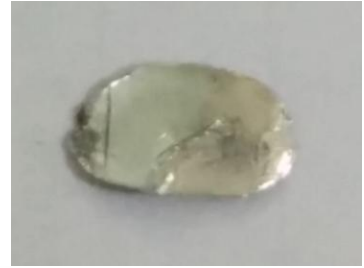

Fig. 2(a)

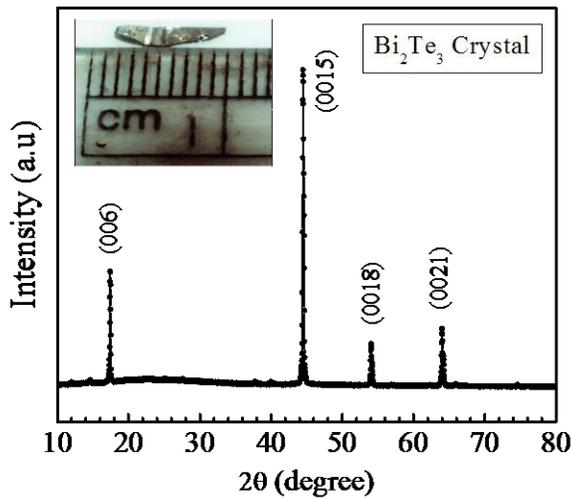

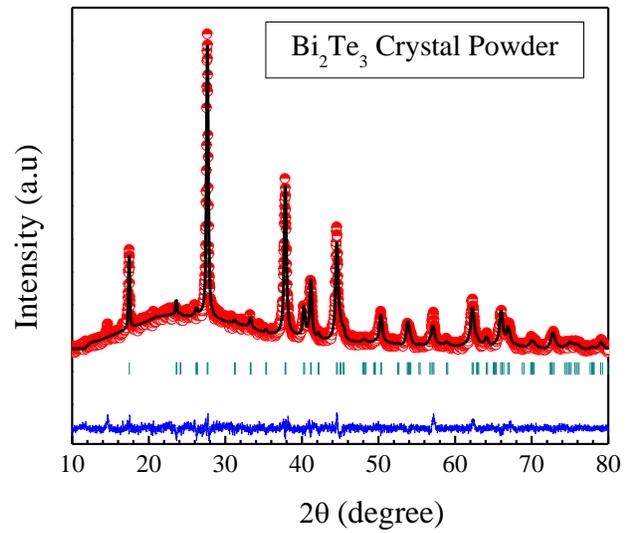

Fig. 2(c)

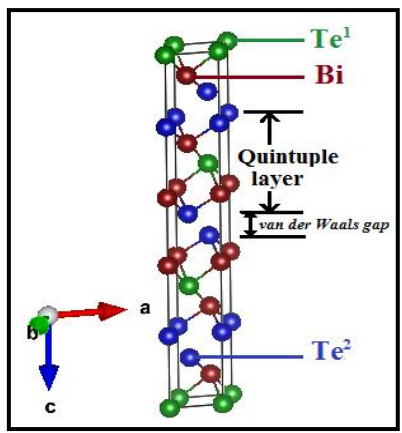

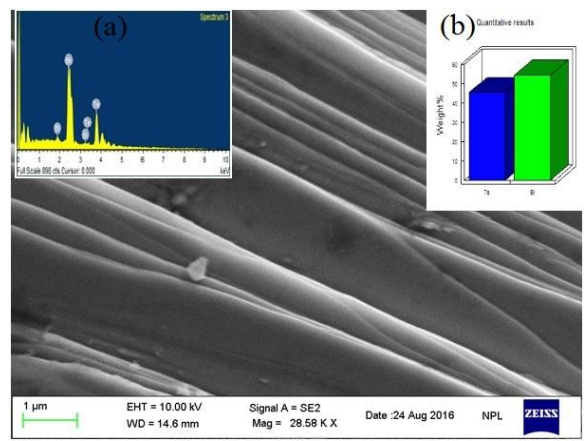

Fig. (4)

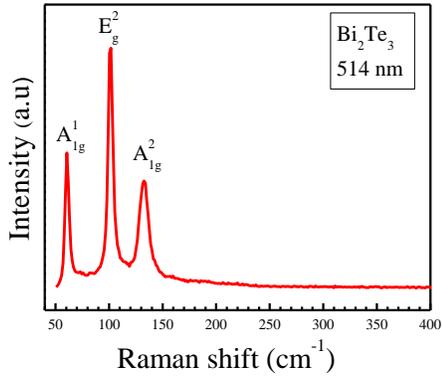

Fig. (5a)

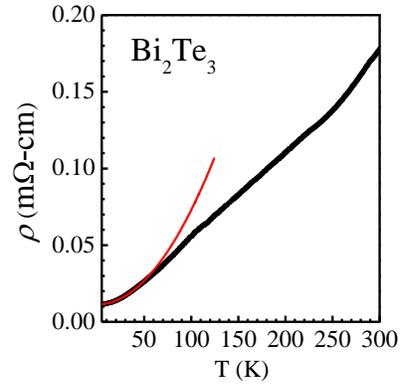

Fig. 5(b)

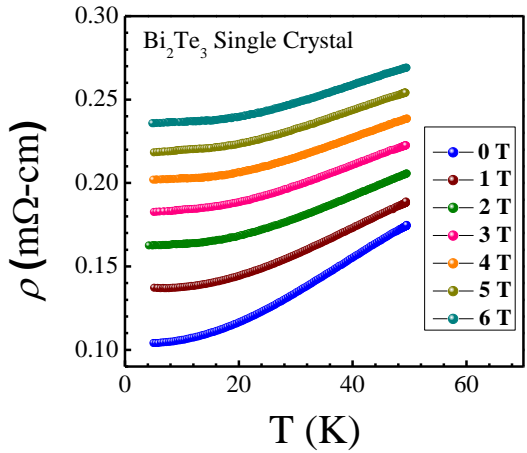

Fig. 5(c)

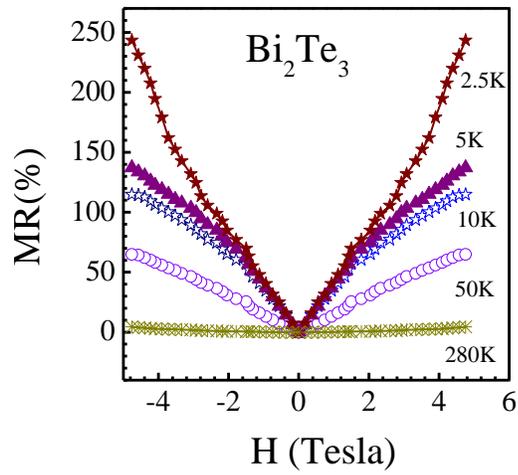

Fig. 5(d)

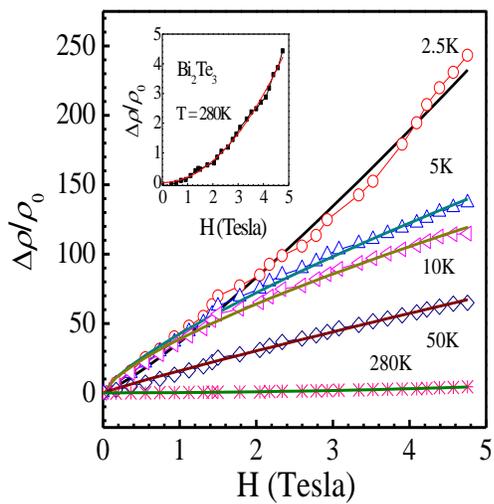

Fig. 5(e)

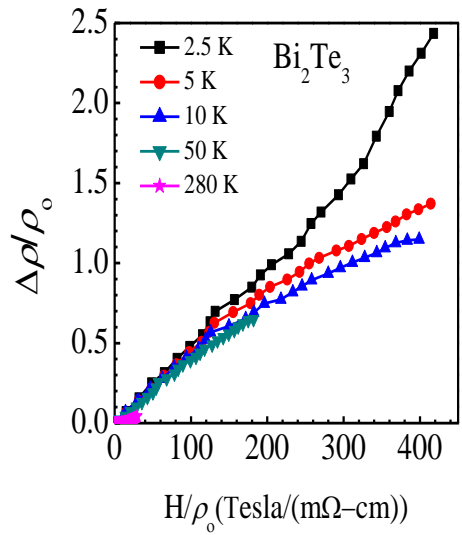